\documentclass[11pt]{article}
\pdfoutput=1
\usepackage[centertags]{amsmath}
\usepackage[square, comma, sort&compress,numbers]{natbib}
\usepackage{array,multirow}
\numberwithin{equation}{section}
\usepackage{amssymb}
\usepackage{graphicx}
\usepackage{color}

\usepackage{epsfig}
\usepackage{bbold}

\def\be{\begin{equation}}
\def\ee{\end{equation}}
\def\ba{\begin{array}}
\def\ea{\end{array}}

\def\dps{\displaystyle}

\def\tr{{\rm Tr}}

\def\1{\tilde{1}}
\def\2{\tilde{2}}
\def\3{\tilde{3}}

\newdimen\tableauside\tableauside=1.0ex
\newdimen\tableaurule\tableaurule=0.4pt
\newdimen\tableaustep
\def\phantomhrule#1{\hbox{\vbox to0pt{\hrule height\tableaurule
width#1\vss}}}
\def\phantomvrule#1{\vbox{\hbox to0pt{\vrule width\tableaurule
height#1\hss}}}
\def\sqr{\vbox{%
  \phantomhrule\tableaustep

\hbox{\phantomvrule\tableaustep\kern\tableaustep\phantomvrule\tableaustep}%
  \hbox{\vbox{\phantomhrule\tableauside}\kern-\tableaurule}}}
\def\squares#1{\hbox{\count0=#1\noindent\loop\sqr
  \advance\count0 by-1 \ifnum\count0>0\repeat}}
\def\tableau#1{\vcenter{\offinterlineskip
  \tableaustep=\tableauside\advance\tableaustep by-\tableaurule
  \kern\normallineskip\hbox
    {\kern\normallineskip\vbox
      {\gettableau#1 0 }%
     \kern\normallineskip\kern\tableaurule}%
  \kern\normallineskip\kern\tableaurule}}
\def\gettableau#1 {\ifnum#1=0\let\next=\null\else
  \squares{#1}\let\next=\gettableau\fi\next}

\newtheorem{prop}{Proposition}[section]
\newtheorem{lemma}[prop]{Lemma}

\newcommand{\bref}[1]{\textbf{\ref{#1}}}

\def\cA{\mathcal{A}}

\def\cF{\mathcal{F}}

\def\cH{\mathcal{H}}

\def\cL{\mathcal{L}}

\def\cO{\mathcal{O}}

\def\cV{\mathcal{V}}

\numberwithin{equation}{section} \makeatletter
\@addtoreset{equation}{section}

\def\ads{AdS_{3}}
\def\be{\begin{equation}}
\def\ee{\end{equation}}
\def\ba{\begin{array}}
\def\ea{\end{array}}

\def\dps{\displaystyle}

\def\ba{\begin{array}}
\def\ea{\end{array}}

\def\dps{\displaystyle}

\def\cl{c\to\infty}

\def\de{\Delta}
\def\f{\psi}

\usepackage{jheppub}

\makeatletter
\def\@fpheader{\vspace{-.1cm}}
\makeatother

\title{Various semiclassical limits of torus conformal blocks}

\author[a,b]{Konstantin\ Alkalaev,}
\author[c]{Roman\ Geiko,}
\author[a,b,d]{Vladimir\ Rappoport}

\affiliation[a]{I.E. Tamm Department of Theoretical Physics, \\P.N. Lebedev Physical
Institute,\\ Leninsky ave. 53, 119991 Moscow, Russia}
\affiliation[b]{Department of General and Applied Physics, \\
Moscow Institute of Physics and Technology, \\
7 Institutskiy per., Dolgoprudnyi, 141700 Moscow region, Russia}
\affiliation[c]{Mathematics Department,\\
National Research University Higher School of Economics,  \\
Usacheva str. 6, 119048 Moscow, Russia}
\affiliation[d]{Department of Quantum Physics, \\
Institute for Information Transmission Problems, \\
Bolshoy Karetny per. 19, 127994 Moscow, Russia}
\emailAdd{alkalaev@lpi.ru}
\emailAdd{rvgeyko@edu.hse.ru}
\emailAdd{belavin@lpi.ru}
\abstract{
We study four types of one-point torus blocks arising in the large central charge regime. According to different limits of conformal dimensions we distinguish between the global block, the light block, the heavy-light block, and the linearized classical block. We show that they are not independent and connected to each other by various links. We find that the global, light,  and heavy-light blocks correspond to three different contractions of the Virasoro algebra.  Also, we formulate the $c$-recursive representation of the one-point torus blocks which is relevant in the semiclassical approximation.
}
\preprint{FIAN-TD-2016-26}
\arxivnumber{}

\begin{document}

\maketitle
\flushbottom

\section{Introduction}

The large central charge CFTs have been receiving  much attention recently in the context of the holographic duality. The Brown-Henneaux relation $c \sim 1/G_N$, where $c$ is the central charge and $G_N$ is the gravitational constant  suggests that the large-$c$ regime corresponds to the semiclassical approximation in the gravity  path integral calculations \cite{Brown:1986nw}. This brings to light many issues regarding studying the leading saddle point gravity description of the large-$c$  CFT and finding the subleading $1/c$ corrections (see, e.g., \cite{Hartman:2013mia,Fitzpatrick:2015zha,Hijano:2015rla,Alkalaev:2015wia,Hijano:2015qja,Banerjee:2016qca,Chen:2016dfb,Alkalaev:2016rjl,Fitzpatrick:2016mjq,Hulik:2016ifr} and references therein).

In this paper we study one-point conformal blocks in various large-$c$ regimes. We distinguish between different large-$c$ torus blocks: the global, light, heavy-light, and classical torus blocks. Three of them are known in the spherical case. We add another type of torus block to this list: the light block. We show that all of them are related by a particular chain of connections.

We shortly  remind the definition of one-point torus conformal block (see, e.g.,
\cite{Cardy:1986ie, Fateev:2009me,Poghossian:2009mk,Hadasz:2009db,Menotti:2010en}). Let $q = e^{2\pi i \tau}$ parameterize the torus  with the complex modulus
$\tau$. The conformal block $\cV(\Delta, \tilde\Delta, c|q)$ of the primary operator  $\phi_{\Delta}(z)$ with the conformal dimension $\Delta$ and the exchanged operator with the conformal dimension $\tilde \Delta$ is given by
\be
\label{Virtor}
q^{c/24-\tilde\Delta}\,\cV(\Delta, \tilde\Delta, c|q) = \sum_{n = 0}^{\infty} q^n \cV_{n}(\Delta, \tilde\Delta, c) =
1+\Big[1+\frac{(\Delta - 1)\Delta}{2\tilde \Delta}\Big] q + \ldots \;,
\ee
where the expansion coefficients are defined as
\be
\label{1ptcoef}
\cV_{n}(\Delta, \tilde\Delta, c) = \frac{1}{\langle \tilde \Delta| \phi_{\Delta}(z) | \tilde \Delta\rangle }\sum_{n = |M|=|N|} B^{M|N}\,
\langle \tilde \Delta, M| \phi_{\Delta}(z) |N, \tilde \Delta\rangle \;,
\ee
where $|\tilde \Delta, M \rangle = L_{-m_1}^{i_1} .... L_{-m_k}^{i_k} |\tilde \Delta\rangle $ are descendant vectors in the Verma module generated from the primary state $|\tilde \Delta\rangle$. Here, $M$ labels basis monomials, $|M| = i_1m_1+ \ldots +i_k m_k$ denotes the sum of the Virasoro generator indices. The matrix $B^{M|N}$ is the inverse of the Gram matrix
$B_{M|N} = {\langle \tilde \Delta, M |N, \tilde \Delta\rangle}$. Note that the block  does not depend on $z$ by virtue of the Ward identity.

We consider the large-$c$ expansions of the block function \eqref{Virtor}. All associated conformal blocks are referred to as {\it semiclassical}. There are different semiclassical blocks because  \eqref{Virtor} also depends  on conformal dimensions $\Delta$ and $\tilde\Delta$ which can be scaled in terms of the central charge. We distinguish between the following four types.
\begin{itemize}
\item $\Delta$ and $\tilde\Delta$ are fixed in the limit $\cl$. Depending on the particular $1/c \to 0 $ contraction of the Virasoro algebra we distinguish between the {\it global}  and  {\it light} blocks. The light block is the leading asymptotic in the $c$-recursive representation of the torus block \eqref{Virtor}.

\item $\Delta$ and $\tilde\Delta/c$ are fixed in the limit $\cl$. In this case we obtain  the {\it heavy-light} block. It follows that $\tilde \Delta \gg 1$ and, therefore, $\Delta/\tilde \Delta \ll 1$. Remarkably, the opposite regime of heavy external operator and light exchanged operator  does not exist for the one-point torus block, cf. the first coefficient in \eqref{Virtor}. It follows that the concept of vacuum block approximation used to calculate conformal blocks on the sphere  \cite{Hartman:2013mia,Fitzpatrick:2014vua,Hijano:2015rla,Alkalaev:2015wia,Alkalaev:2015lca,Datta:2014zpa} (see  \cite{Rangamani:2016dms} for review) is not directly  applicable to one-point torus blocks because the exchanged operators wrap around  non-vanishing cycle and, therefore, cannot be set to $\mathbb{1}$.

\item $\Delta/c$ and $\tilde\Delta/c$ stay fixed as $\cl$. The resulting asymptotic expansion of the block function \eqref{Virtor}  is the exponential of the {\it classical} conformal block. The  linearized version of the classical torus block is obtained when the ratio $\Delta/\tilde \Delta$ is small. Note that the linearized classical block admits the holographic interpretation in the thermal $\ads$ space \cite{Alkalaev:2016ptm}. In the present paper we find the concise  integral representation of the linearized classical block.

\end{itemize}

The paper is organized as  follows. Each of the following Sections from \bref{sec:glob}  to \bref{sec:classical}  considers a particular  type of semiclassical torus blocks listed above and establishes  their mutual relations. In Section \bref{sec:con} we discuss our conclusions and further perspectives. Appendix \bref{sec:C} collects first four block coefficients, Appendix \bref{sec:D} shows the relation between the global and light blocks, Appendix \bref{sec:A} considers the Kac determinant in the $\cl$ limit, Appendix \bref{sec:rec} describes the $c$-recursive representation of the one-point torus block.

\section{Global torus block}
\label{sec:glob}

The Virasoro symmetry algebra $Vir$ on any Riemann surface contains a finite-dimensional subalgebra $sl(2, \mathbb{C}) \subset Vir$. It follows that we can always restrict the general Virasoro block to that of the $sl(2, \mathbb{C})$ subalgebra. The resulting global block is a simpler function which depends on the conformal dimensions and not on the central charge. Note that the general conformal block of higher dimensional CFT being restricted to two dimensions yields the global block.

In the torus case, the general formula \eqref{1ptcoef} simplifies to yield the global torus block
\be
\label{glob}
\cF(\Delta, \tilde\Delta|q)=\sum_{n=0}^\infty q^n \cF_{n}(\Delta, \tilde\Delta)\;,
\qquad
\cF_{n}(\Delta, \tilde\Delta) = \frac{1}{\langle \tilde \Delta| \phi_{\Delta}(z) | \tilde \Delta\rangle }\frac{\langle \tilde \Delta|
L_{1}^n \phi_{\Delta}(z) L^n_{-1}|\tilde \Delta\rangle}{\langle \tilde \Delta|L^n_1L^n_{-1}|\tilde \Delta\rangle}\;,
\ee
where $L_{0, \pm1}$ are $sl(2, \mathbb{C})$ basis elements.  The coefficients can be found explicitly\footnote{A detailed derivation of  $sl(2, \mathbb{C})$ matrix elements can be found in \cite{Alkalaev:2015fbw}.}
\be
\label{1ptcoefGLOBAL}
\cF_{n}(\Delta, \tilde\Delta)  = \sum_{p=0}^n
\frac{n!}{(n-p)!(p!)^2}\frac{\Gamma(2\tilde \Delta)}{\Gamma(2\tilde \Delta+p)}\frac{\Gamma(\Delta+p)}{\Gamma(\Delta-p)}\;.
\ee
The global block coefficients are related to the hypergeometric function \cite{Hadasz:2009db}
\be
\label{hadasz}
\ba{l}
\dps
\cF(\Delta, \tilde\Delta|q) =  (1-q)^{-1} \;{}_2 F_{1}(\Delta, 1-\Delta, 2\tilde \Delta\, |\, \frac{q}{q-1})  =
\\
\\
\dps
 = 1+\Big[1+\frac{(\Delta - 1)\Delta}{2\tilde \Delta}\Big] q + \Big[1+\frac{(\Delta -1) \Delta }{2 \tilde\Delta }+\frac{(\Delta -1)
 \Delta  (\Delta ^2-\Delta +4 \tilde\Delta)}{4 \tilde \Delta  (2 \tilde\Delta +1)}\Big]q^2+ ... \;.
\ea
\ee

The $\tilde \Delta \to \infty$ global block coefficients can be equivalently described as the $\Delta \to 0$ limiting case. In this case we arrive at the zero-point  block coinciding with the $sl(2, \mathbb{C})$ character,
\be
\label{sl2char}
\cF(\Delta, \tilde\Delta|q)\Big|_{\tilde\Delta\to \infty} =  \frac{1}{1-q} = 1 + q +q^2+q^3+ \ldots\;,
\ee
which indicates  that there is just one state on each level of the corresponding  $sl(2, \mathbb{C})$ module.

\subsection{Large conformal dimensions}
\label{sec:large}

The expression \eqref{hadasz} suggests that the global block function satisfies the Gauss hypergeometric equation. To this end, using the Pfaff transformation the global torus block \eqref{hadasz} can be cast into the form \footnote{Recall that the 4-point global  block on the sphere is also given by the hypergeometric function. It is tempting to speculate that the recently established correspondence between the
one-point torus block and the four-point spherical block \cite{Fateev:2009me,Poghossian:2009mk,Hadasz:2009db} is exact in the $sl(2, \mathbb{C})$ case once the conformal dimensions satisfy appropriate linear relations, while $q$ and $z$ are related in
the standard way.}
\be
\label{pfaff}
\cF(\Delta, \tilde\Delta|q) =  (1-q)^{-\Delta} \;{}_2 F_{1}(2\tilde \Delta - \Delta, 1-\Delta, 2\tilde \Delta\, |\, q)\;.
\ee
Then, we find the second order differential equation 
\be
\label{difur}
\cF^{''}+ 2\left(\frac{\tilde \Delta}{q}-\frac{1}{1-q}\right)\cF^{'}
-\left(\frac{\Delta(\Delta-1)}{q(1-q)^2}+\frac{2\tilde\Delta}{q(1-q)}\right)\cF = 0\;,
\ee
which has three singular points $0,1$ and $\infty$. One of two independent solutions is given by \eqref{pfaff}. We note that global blocks on the sphere satisfy the analogous second order differential equations following from imposing the $sl(2, \mathbb{C})$ Casimir eigenvalue equations in the exchanged channels \cite{Dolan:2011dv,Alkalaev:2015fbw}. The torus case is more intricate because  the global symmetry here is $u(1)\oplus u(1) \subset sl(2, \mathbb{C})$ so that the term global is somewhat misleading. On the other hand, the equation \eqref{difur} should be realized as the torus Casimir channel condition.

We consider the regime where dimensions $\Delta$ and $\tilde \Delta$ tend to infinity in a coherent manner. Namely, we introduce the scale parameter $\kappa$ as
\be
\label{kappa}
\Delta = \kappa\,  \sigma\;, \qquad \tilde\Delta  = \kappa \,\tilde \sigma\;,
\ee
where  $\sigma$ and $\tilde\sigma$ are rescaled conformal dimensions.  Below we study the large-$\kappa$ asymptotic behavior of \eqref{pfaff} given that  $\sigma$ and $\tilde \sigma$ are fixed. We prove the following
\begin{prop}
\label{propglob}
The large-$k$ asymptotic expansion of the global block function is exponentiated
\be
\label{globexp}
\cF(\Delta, \tilde \Delta|q) = \exp\left[\,\kappa\, g_0(\sigma, \tilde \sigma|q)\right]\;,
\qquad\;
\kappa \to \infty\;,
\ee
where $g_0(\sigma, \tilde \sigma|q)$ is given by the formal power series
\be
\label{gofinn}
g_0(\sigma, \tilde \sigma|q) = \tilde \sigma \sum_{n=1}^{\infty} g_{0n}(q) \left(\frac{\sigma}{\tilde\sigma}\right)^{2n}\;,
\qquad
g_{0n}(q) = \frac{q^n}{\varkappa_n} \,\frac{q^{n-1}+ \gamma_{n,n-2} q^{n-2}+...+ \gamma_{n,0}}{(1-q)^{2n-1}}\;,
\ee
where $\gamma_{n,i} = (-1)^{n-i+1}\binom{2n-1}{n+i}$, $i = 0,1,...\,,n-1$ are binomial coefficients, and $\varkappa_n =\frac{4^n (2n-1)n}{2}$.

\end{prop}
The proof directly follows from the  theory of differential equations with a large parameter. In general, the large-$\kappa$ asymptotic expansion takes the form
\be
\label{expansionF}
\cF(\Delta, \tilde \Delta|q)=\exp\left[\sum_{n=0}^\infty \kappa^{\lambda - n} g_{n}(\sigma, \tilde \sigma|q)\right]\;,
\ee
where $\lambda$ is a  positive integer. The expansion coefficients $g_{n}(\sigma, \tilde \sigma|q)$ can be explicitly found by substituting \eqref{expansionF} into the original differential equation \eqref{difur}. The resulting recurrence equations fix $\lambda=1$, while the first coefficient $g_{0}(\sigma, \tilde \sigma|q)$ satisfies the first order differential equation
\be
\label{quadra}
(g^{'}_{0}(\sigma, \tilde \sigma|q))^2+\frac{2\tilde\sigma}{q}\,g^{'}_{0}(\sigma, \tilde \sigma|q)-\frac{\sigma^2}{q(1-q)^2}=0\;,
\ee
where the prime denotes the $q$-derivative. All other coefficients in \eqref{expansionF} are expressed in terms of $g_{0}(\sigma, \tilde \sigma|q)$ and its derivatives. The quadratic equation \eqref{quadra} yields two values of the derivative of which we choose the solution which is perturbatively regular at $q=0$. We integrate to obtain
\be
\label{gint}
g_{0}(\sigma, \tilde \sigma|q)=\int_0^q dx\left( -\frac{\tilde\sigma}{x}+\sqrt{\frac{\tilde\sigma^2}{x^2}+\frac{\sigma^2}{x(1-x)^2}}\,\right)\;.
\ee
The integration constant is chosen so that $\cF(\Delta, \tilde \Delta|0) = 1$, or, equivalently, $g_{0}(\sigma, \tilde \sigma|0)=0$. We first decompose the right-hand side of \eqref{gint} with respect to the small parameter $\delta = \sigma/\tilde\sigma$. Expanding the integrand and integrating the resulting expression we arrive at \eqref{gofinn}.

\section{Light  torus block}

In general, the light block can be  defined as the $\cl$ limit of the quantum conformal block. For example, the limiting
$4$-point function on the  sphere is known to be the hypergeometric function \cite{Zamolodchikov:1987ie}. In the 5-point case this is the Horn hypergeometric series of two variables \cite{Alkalaev:2015fbw}, the $n$-point case was studied in \cite{Bhatta:2016hpz}.  The light blocks are also known in the $W_N$ case \cite{Fateev:2011qa,Poghosyan:2016lya}. The light and global blocks on the sphere  are identical.

We expand conformal blocks given that $\Delta = \cO(c^0)$ and $\tilde \Delta = \cO(c^0)$ as $\cl$. In this regime the quantum torus block \eqref{Virtor} is represented as
\be
\label{lightb}
q^{c/24-\tilde\Delta}\,\cV(\Delta, \tilde\Delta, c|q) = \cL(\Delta, \tilde\Delta|q) + \cO(c^{-1}) \;,
\ee
where the leading term is the  one-point  light torus block. It turns out that global and light blocks are different on a
torus. The light block is a formal power series
\be
\label{lbcoef}
\ba{l}
\dps
\cL(\Delta, \tilde \Delta|q) = \sum_{n=0}^\infty \cL_n(\Delta, \tilde \Delta) \, q^n\; =
\\
\\
\dps
\hspace{1cm}=1+\Big[1+\frac{(\Delta - 1)\Delta}{2\tilde \Delta}\Big]q
+\Big[2+\frac{(\Delta -1) \Delta }{2 \tilde\Delta }+\frac{(\Delta -1)
 \Delta  (\Delta ^2-\Delta +4 \tilde\Delta)}{4 \tilde \Delta  (2 \tilde\Delta +1)}\Big]q^2 + \ldots\;,
\ea
\ee
cf. \eqref{hadasz}. There is the following
\begin{prop}
\label{gl}
The global and light blocks are related as
\be
\label{lightglob}
\cL(\Delta, \tilde \Delta|q) = \frac{1-q}{\varphi(q)}\,\cF(\Delta, \tilde \Delta|q) \;,
\ee
where $\varphi(q) = \dps\prod_{n=1}^\infty (1-q^n)$ is the Euler function \eqref{1generating}.
\end{prop}
The proof is given in Appendix \bref{sec:D}. The component form of \eqref{lightglob} reads
\be
\label{LvsF}
\cL_{n}(\Delta, \tilde\Delta)=\sum_{k=0}^{n}p_1(n-k)\mathcal{F}_k(\Delta, \tilde\Delta)\;,
\ee
where $p_1(n-k)$ is the number of partitions of $n-k$ which does not contain $1$ as a part, with the convention that $p_1(0) = 1$. Equivalently, $\cL_n  = \left[\cF_n + \cF_{n-2} + \cF_{n-3}+p_1(4)\cF_{n-4}+ ... \,\right]$.

In brief, to prove Proposition \bref{gl} we analyze the primary operator matrix and the Gram matrix elements which are polynomial functions of the central charge. \footnote{See the analogous discussion of the global blocks on the sphere using the projector technique \cite{Fitzpatrick:2014vua}. Also, see the large $\tilde \Delta $ asymptotics analysis in the torus case  \cite{Kraus:2016nwo}.} Note that the $n$-th level block coefficients can be represented as a trace of the $[n\times n]$ matrix product, see \eqref{1ptcoef}. Thus, the block coefficients are rational functions of $c$. In Appendix \bref{sec:D} we show that only diagonal elements contribute in the $\cl$ limit. On the other hand, the diagonal elements are exactly the global block coefficients weighted with the number of the $sl(2, \mathbb{C})$ basis elements $L_{-1}$ on a given level. This explains the weight factor in \eqref{lightglob}.

\subsection{Virasoro algebra contractions }
\label{sec:contracted}

We consider the In\"{o}nu-Wigner  contraction \cite{Inonu:1953sp} for the Virasoro algebra, where the deformation  parameter is the inverse central charge. In the $\cl$ limit, there are two different types of contractions underlying the global and light torus blocks. One of contractions gives rise to the Heisenberg algebra that explains the weight prefactor in \eqref{lightglob}.

We recall that the Virasoro algebra commutation relations are given by
\be
\label{virasoro}
[L_m, L_n] = (m-n)L_{m+n}+ \frac{c}{12} m(m^2-1)\delta_{m+n,0}\;, \qquad m,n \in \mathbb{Z}\;,
\ee
while primary operators transform as
\be
\label{primary}
[L_m, \phi_\Delta] = z^m(z\partial_z +(m+1)\Delta)\phi_{\Delta}\;.
\ee
The  rescaled Virasoro generators are $L_m \to c^{-\gamma(m)} L_m$, where $\gamma(m)$ is some function of the label $m \in \mathbb{Z}$. There are two natural cases: (A) $\gamma(|m|\leq 1) = 0$ and $\gamma(|m|\geq 2) = 1$; (B)  $\gamma(|m|\leq 1) = 0$ and $\gamma(|m|\geq 2) = 1/2$. We denote
\be
L_{0, \pm 1} \to l_{0,\pm 1} =  L_{0, \pm 1} \;, \qquad\quad L_{m} \to a_m =  L_{m}/c^\gamma\;, \quad  |m|\geq 2\;,
\ee
where $\gamma = 1$ and $\gamma = 1/2$ correspond to the cases (A) and (B), respectively.

The transformation law \eqref{primary} is also rescaled. In the limit $\cl$, keeping the conformal dimension $\Delta$ finite we find that in both cases (A) and (B) the primary operator  transforms as
\be
\label{slphi}
[l_m, \phi_\Delta]  = z^m(z\partial_z +(m+1)\Delta)\phi_{\Delta}\;,
\qquad\;\;
[a_m, \phi_\Delta]  = 0\;.
\ee
It follows that  $\phi_\Delta$ are $sl(2)$ conformal operators and $a_m-$singlets.

In the case (A), the contracted Virasoro algebra splits into  $sl(2)$ algebra and the infinite-dimensional Abelian algebra $\cA$,
\be
\label{slA}
\ba{c}
\dps
[l_{m}, l_n]  = (m-n)l_{m+n}\;,
\qquad\qquad [a_m, a_n] = 0\;,
\vspace{3mm}
\\
\dps
[l_m, a_n] = (m-n)a_{m+n}\;, \qquad |m+n|\geq 2\;;
\qquad
[l_m, a_n] = 0\;, \qquad |m+n|\leq 1\;.
\ea
\ee
The contracted algebra is the semidirect sum $Vir_A = sl(2)\ltimes \cA$, while the $\pm$ branches of $\cA$ are lowest weight $sl(2)$--modules.

In the case (B), the contracted Virasoro algebra  splits into $sl(2)$ algebra and the infinite-dimensional Heisenberg algebra $\cH$,
\be
\label{slH}
\ba{c}
\dps
[l_{m}, l_n]  = (m-n)l_{m+n}\;,
\qquad\qquad
[a_m, a_n] = \frac{m(m^2-1)}{12}\delta_{m+n,0}\;,
\vspace{3mm}
\\
\dps
[l_m, a_n] = (m-n)a_{m+n}\;, \quad |m+n|\geq 2\;;
\qquad
[l_m, a_n] = 0\;, \quad |m+n|\leq 1\;.

\ea
\ee
The contracted algebra in this case is the semidirect sum $Vir_B = sl(2)\ltimes \cH$, while the $\pm$ branches of $\cH$ are lowest weight $sl(2)$--modules. Comparing \eqref{slA} and \eqref{slH} we see that the only difference is the central term in \eqref{slH}, which refers to that the Heisenberg algebra is the centrally extended Abelian algebra.

We argue that the contracted algebras $Vir_A$ and $Vir_B$ underlie the global and light torus blocks, respectively. Let the basis monomials be represented as
\be
\label{ALsbasis}
|M, \tilde \Delta\rangle  =  a_{\bar M} l_{-1}^s|\tilde \Delta\rangle\;, \qquad |M| = s+ |\bar M|\;,
\ee
where we denoted $a_{\bar M} = a_{-m_1}^{i_1} \ldots  a_{-m_k}^{i_k}$ and $i_1m_1 + \dots +i_k m_k = |\bar M|$. The standard conjugation rules $a_{-m} = (a_m)^\dagger$ are assumed.

The global torus block corresponds to the case (A). We compute the block coefficients \eqref{1ptcoef}, where the basis monomials $|\tilde \Delta, M \rangle$ are elements of the contracted algebra $Vir_A$ module, while the primary operator satisfies \eqref{slphi}. The Abelian factor $\cA$ is trivially realized and the only states  contributing to the block coefficients are those with $\bar M=0$. It follows that the $n$-th level block coefficient contains the maximum number of $l_{\pm 1}$ generators, $n = s$ so that  the resulting expression is just the the  global  block \eqref{glob}. We note that truncating the Virasoro algebra to the $sl(2, \mathbb{C})$ subalgebra of Section \bref{sec:glob} is equivalent to considering the contracted $Vir_A$ with trivially realized Abelian factor.

The  light torus block corresponds to the case (B). To compute the $Vir_B$  block we observe  that the Gram matrix  and the primary field matrix  are diagonal, $\langle \tilde \Delta, M|N, \tilde \Delta\rangle = \delta_{MN}\alpha_{M}\langle \tilde \Delta|l_1^{s} l_{-1}^s |\tilde \Delta\rangle$ and $\langle \tilde \Delta, M|\phi_{\Delta}|N, \tilde \Delta\rangle = \delta_{MN}\beta_{M}\langle \tilde \Delta|l_1^{s} \phi_{\Delta} l_{-1}^s |1\tilde \Delta\rangle$, where $\alpha_M = \beta_N$ are some non-zero constants, the bra and ket vectors are given in the basis  \eqref{ALsbasis}.  Using that $\cH$ is the ideal and $a_m |\tilde \Delta\rangle = 0$ we find the relation  $[a_m, l_{-1}^s]\,|\tilde \Delta\rangle = 0$ used to show the first equality above. The second equality follows from the first one  and the commutator \eqref{slphi}. It follows that the number $s$ of $l_m$ generators in the matrix elements on the $n$-th level varies from  $0$ to $n$.

We see that, contrary to the case (A), the Heisenberg factor $\cH$ is non-trivially realized and, therefore, the global block associated to the $sl(2)$ factor and the block of the full $Vir_B = sl(2)\ltimes \cH$ become related. When computing the expansion coefficients \eqref{1ptcoef} on the $n$-th level the infinite-dimensional Heisenberg algebra $\cH$ can be effectively reduced  to the finite-dimensional Heisenberg algebra $\cH_{p(n)+1}$ with $p(n)+1$ basis elements, where $k(n)$ is a number of partitions of $n$.  Then, using the diagonal matrix  relations we find that the corresponding conformal block is given by the sum of the global block coefficients $\cF_i$, $i = n, n-2, n-3,..., 1,0$ given by \eqref{glob} with weights equal to the number of $l_{-1}$ generators in the basis monomials on the $n$-th level, cf. \eqref{LvsF}. Indeed, the basis \eqref{ALsbasis} has a natural grading which is the number of $l_{-1}$ generators so that we can organize all basis elements according to the $s$-grading. Then, the first few elements on the $n$-th level are: $l_{-1}^{n}|\tilde \Delta\rangle$, $a_{-2}l_{-1}^{n-2}|\tilde \Delta\rangle$, $a_{-3}l_{-1}^{n-3}|\tilde \Delta\rangle$, etc.  Thus, we find that the $Vir_B = sl(2)\ltimes \cH$ block  indeed is  identified with the light  block expressed in terms of the global  block \eqref{lightglob}.

\subsection{Torus $c$-recurrence}

The light block can be considered as  the leading $\cO(c^0)$ term of the so-called $c$-recursive  representation of the quantum
conformal block, where sub-leading $\cO(1/c)$ terms are simple poles in $c$ \cite{Zamolodchikov:1987ie}. In the torus case, the $\cl$  limit of the original block function is identified with the light block \eqref{lightb}. The recursive relations are given by
\be
\label{torusrec}
    \cV_n(\Delta,\tilde\Delta, c ) = \cL_n(\Delta, \tilde\Delta) + \sum_{r\ge2,s\ge1}^{rs\le n}\bigg(-\frac{\partial c_{r,s}}{\partial
    \tilde \Delta}\bigg)\frac{A_{rs}(\tilde \Delta) P_{rs}(\Delta,\tilde \Delta)}{c-c_{rs}(\tilde\Delta)}\cV_{n-rs}(\Delta,
    \tilde\Delta+rs, c)\;,
 \ee
where the   central charge values $c_{r,s}(\tilde \Delta)$ associated to the degenerate dimensions  and the functions $A_{rs}(\tilde \Delta)$, $P_{rs}(\Delta,\tilde \Delta)$ are given in Appendix \bref{sec:rec}. Schematically, the recursive representation defines the $n$-th block coefficient as $\cV_n = \cL_n + \eta_{n-2}\cV_{n-2}+ \eta_{n-3}\cV_{n-3}+ ... + \eta_1 \cV_1 + \eta_0 \cV_0$, where $\eta$-s are read off from \eqref{torusrec}.

\section{Heavy-light torus block}
\label{sec:HL}

The heavy-light limit of conformal blocks assumes that some dimensions scale linearly with the central charge,  while the others remain fixed \cite{Fitzpatrick:2015zha,Fitzpatrick:2014vua}. Thus, in the large-$c$ regime we distinguish between heavy and light operators.

The torus one-point block depends on two conformal dimensions, $\Delta$ and $\tilde \Delta$. In this case, the heavy-light limit is defined by $\Delta = \cO(c^0)$ and $\tilde \Delta = \cO(c^1)$ as $\cl$.
Note that contrary to the spherical blocks the external operator is light and the exchanged operator is heavy. In the opposite regime the torus block  does not exit because the block function diverges. It follows that there is just one heavy-light torus block defined as
\be
\label{hlb}
\cH(\Delta, \tilde \epsilon\,|q) =  \lim_{\substack{ \cl\\ \Delta, \tilde\epsilon -\text{fixed}}}q^{c/24-\tilde\Delta}\, \cV(\Delta, c \tilde \epsilon, c\,|q)\;,
\ee
where $\tilde \epsilon = \tilde \Delta/c$ is the classical exchanged dimension. The block coefficients calculated to high enough orders are given by
\be
\label{hlbc}
\cH(\Delta, \tilde \epsilon\,|q) = 1+ q+ 2q^2 +3q^3 + 5q^4+ \ldots = \frac{1}{\varphi(q)}\;,
\ee
where $\varphi(q)$ is the Euler function, cf. \eqref{lightglob}. Here, the last equality is the conjecture based on the middle terms following  from explicit calculations of the first few levels. Below we argue  that the heavy-light block is indeed the inverse Euler function. In this case, the heavy-light block is just the zero-point block, or, equivalently, the Virasoro character. Note that the heavy-light block  turns out to be independent of the conformal dimensions, $\cH(\Delta, \tilde \epsilon\,|q)\equiv \cH(q)$.

\vspace{-2mm}

\begin{prop}
\label{heavy}
The heavy-light  block is the limiting case of the light block
\be
\label{lightvshl}
\cH(q) = \lim_{\tilde \Delta \to \infty} \cL(\Delta, \tilde \Delta|q)\;.
\ee
\end{prop}
\noindent The proof directly follows from the global/light block relation  \eqref{lightglob} and the large-$\tilde \Delta$ asymptotic of the global block \eqref{sl2char}.

The relation \eqref{lightvshl} is analogous to that between global and heavy-light blocks on the sphere originally found in the 4-point case \cite{Fitzpatrick:2015zha} and further developed in the higher-point case \cite{Alkalaev:2015fbw}. Recall that two heavy external operators  of the spherical blocks can be  generated by  the singular conformal map $z \to z^\alpha$, where $\alpha$ is expressed in terms of the heavy dimensions. Then the heavy-light block is just the global block with the light operator insertions evaluated in the new coordinates supplemented with the corresponding Jacobian factors. From the holographic perspective, the conformal map generates heavy insertions on the boundary that corresponds to the conical singularity/BTZ in the bulk. In the torus case, the  external operator is lighter than the exchanged operator and therefore it cannot produce a  singularity like in the spherical case. The corresponding bulk geometry is the thermal AdS space which is the standard AdS with periodic time \cite{Alkalaev:2016ptm}. Thus, there is no additional Jacobian factors and the light torus block which replaces the global block in the spherical case is related to the heavy-light block as in \eqref{lightvshl}.

Similarly to the case of the global and light blocks, the heavy-light block \eqref{hlbc} is related to the particular contracted  Virasoro algebra. Indeed, we rescale the Virasoro generators as
\be
L_{0} \to l_{0} =  L_{0}/c \;, \qquad L_{m} \to l_m =  L_{m}/\sqrt{2 c}\;, \quad  m\neq 0 \;,
\ee
and find that in the limit $\cl$ the contracted Virasoro algebra commutation relations read
\be
\label{cHeis}
[l_m, l_{n}] = m \delta_{m+n,0} \,l_0+\frac{m(m^2-1)}{24}\delta_{m+n,0}\;,\qquad m,n \in \mathbb{Z}\;,
\ee
while the primary operator transformation law is given by
\be
\label{prim}
[l_m, \phi_\Delta(z)] = 0\;, \qquad m \in \mathbb{Z}\;.
\ee

The contracted Virasoro algebra is the infinite-dimensional Heisenberg algebra $\cH$. The commutation relations \eqref{cHeis} can be cast into the standard form by  linearly transforming the basis elements $l_m$.   Note that the original $sl(2)$ subalgebra is contracted \footnote{See also Ref. \cite{Barut:1970qf}.} into the three-dimensional Heisenberg algebra $\cH_3 \subset \cH$. Most importantly,  the external operator is the $\cH$-singlet. It follows that the resulting  block coefficients \eqref{1ptcoef}  are simply  the traces of the  $[p(n)\times p(n)]$ unit matrices,
\be
\cH_n = \tr\, \mathbb{1}_{\,p(n)}\equiv p(n)\;,
\ee
where $p(n)$ is the partition of the level number $n$, cf. \eqref{hlbc} and \eqref{1generating}. The particular form of the corresponding Gram matrix makes no difference.

\section{Linearized classical torus block}
\label{sec:classical}

As opposed to the limiting block functions considered in the previous sections, here we study the large-$c$ asymptotic expansion of the torus block, where all conformal dimensions grow linearly with the central charge $\Delta = \cO(c^1)$ and $\tilde \Delta = \cO(c^1)$. It means that both the external and exchanged operators are heavy.  Decomposing the block function \eqref{Virtor} around $c = \infty$ we arrive at the following Laurent series,
\be
\label{laurent}
\cV(\Delta, \tilde \Delta,c|q)  =  \sum_{n\in \mathbb{N}} \frac{v_n(\epsilon, \tilde \epsilon|q)}{c^n}\;,
\ee
where finite parameters
\be
\label{epsilon}
\epsilon = \frac{\Delta}{c}
\qquad\text{and}\qquad
\tilde\epsilon  = \frac{\tilde \Delta}{c}\;,
\ee
are classical conformal dimensions, and $v_n(\epsilon, \tilde \epsilon|q)$ are formal power series in the modular parameter $q$ with expansion coefficients being rational functions in $\epsilon$ and $\tilde \epsilon$. Note that the expansion \eqref{laurent} is essentially different from the c-recursive representation \eqref{torusrec} because of  the rescaling \eqref{epsilon}.

At large $c$ the principle part of \eqref{laurent} tends to zero. Less obvious is the fact that  the regular part exponentiates  \cite{Zamolodchikov1986}. It follows that the one-point torus block is asymptotically equivalent to
\be
\label{ccb}
\cV(\Delta, \tilde \Delta,c|q) \sim  \exp\big[\,c\,f(\epsilon, \tilde \epsilon|q)\big]\qquad  \text{as}\qquad  \cl\;.
\ee
Here, the function $f(\epsilon, \tilde \epsilon|q)$ is the  classical conformal block \cite{Piatek:2013ifa},
\be
\label{class_block}
f(\epsilon, \tilde \epsilon|q) = (\tilde\epsilon - 1/4) \log q + \sum_{n=1}^\infty q^n \text{f}_n (\epsilon, \tilde \epsilon) = (\tilde\epsilon - 1/4) \log q + \frac{\epsilon^2}{2\tilde\epsilon}\,q+ \ldots \;,
\ee
other lower level coefficients can be found in \cite{Piatek:2013ifa,Alkalaev:2016ptm}. Note that coefficients $\text{f}_n (\epsilon, \tilde \epsilon)$ are combinations of the expansion coefficients of $v_{m\geq 0}(\epsilon, \tilde \epsilon|q)$ in \eqref{laurent}.

The torus one-point linearized classical block is defined by introducing the lightness parameter $\delta = \Delta/\tilde\Delta<1$,
which measures the relative weight of dimensions  at any value of the central charge, $\Delta/\tilde \Delta =\epsilon/\tilde\epsilon$ \cite{Alkalaev:2015wia,Alkalaev:2015fbw}.
Changing from $(\epsilon, \tilde \epsilon)$ to $(\delta,\tilde \epsilon)$ we represent the  classical conformal block  as a double series expansion in $\delta$ and $\tilde\epsilon$,
\be
\label{vovalin1}
f(\delta \tilde\epsilon,\tilde\epsilon|q)=\sum_{n=2}^{\infty} \f_{n}( \tilde\epsilon|q)\, \delta^{n}\;,
\qquad
\text{where}
\qquad
\f_n( \tilde\epsilon|q)=\sum_{m=0}^\infty \f_n^{(m)}(q) \tilde\epsilon^{\,m}\;.
\ee
Keeping terms at most linear in $\tilde \epsilon$ in the original classical block we obtain the linearized version, $f(\epsilon,\tilde{\epsilon}|q) = f^{lin}(\delta,\tilde\epsilon|q)+\mathcal{O}(\tilde\epsilon^2)$, where the lightness parameter is dimensionless \cite{Alkalaev:2016ptm}. The linearized classical block $f^{lin}(\delta,\tilde\epsilon|q)$ is given by the expansion coefficients  $\f_n^{(1)}(q)$ \eqref{vovalin1}. It turns out that $\f_{2k+1}^{(1)}(q)=0$ at $k=1,2,...\,$. Denoting $ \f_{2k}^{(1)}(q)\equiv f_k^{(1)}(q)$ we find the linearized classical torus block
\be
\label{vovalin2}
f^{lin}(\delta,\tilde\epsilon|q)\equiv (\tilde\epsilon - 1/4) \log q + \tilde \epsilon \sum_{n=1}^{\infty} f^{(1)}_{n}(q) \delta^{2n}\;.
\ee
The expansion coefficients in \eqref{vovalin2} can be explicitly calculated at  $n\leq 5$ that allows us to conjecture   the general formula \cite{Alkalaev:2016ptm}
\be
\label{lincoefs}
f^{(1)}_n = \frac{q^n}{\varkappa_n} \,\frac{q^{n-1}+ \gamma_{n,n-2} q^{n-2}+...+ \gamma_{n,0}}{(1-q)^{2n-1}}\;,
\ee
where  $\gamma_{n,i} = (-1)^{n-i+1}\binom{2n-1}{n+i}$, $i = 0,1,...\,,n-1$ are binomial coefficients, and $\varkappa_n$ are some constants  $\varkappa_1 = 2$, $\varkappa_2 = 48$, $\varkappa_3 = 480$, $\varkappa_4 = 3584$, $\varkappa_5 = 23040$, etc. \footnote{Typically, the linearized classical blocks both on the sphere and torus are infinite series in the dimensionless lightness parameter  \cite{Alkalaev:2015wia,Alkalaev:2015fbw,Alkalaev:2016ptm}. An exception to this is the $4$-point linearized classical  block on the sphere which is a linear function in conformal dimensions \cite{Fitzpatrick:2014vua,Hijano:2015rla}.}

Remarkably, the linearized classical torus block is related to the global block at large dimensions. There is the following
\begin{prop}
\label{fkw}
Coefficients $g_{0n}(q)$ of the exponentited global block  \eqref{gofinn} and the linearized classical block coefficients $f^{(1)}_{n}(q)$ \eqref{vovalin2} are equal to each other,
\be
\label{globlin}
f^{(1)}_{n}(q) = g_{0n}(q)\;.
\ee
\end{prop}
\noindent This can be verified  by explicit calculation at $n\leq 5$. Put differently, the relation \eqref{globlin} says that changing from $(\varkappa, \sigma, \tilde \sigma)$ of the global block \eqref{kappa} to $(c, \epsilon, \tilde\epsilon)$ of the classical block \eqref{epsilon} we find that the linearized classical block is exactly equal to the exponential factor of the global block at large dimensions,
\be
\label{g0f}
g_0(\epsilon, \tilde \epsilon|q) = f^{lin}(\epsilon,\tilde\epsilon|q)\;,
\ee
cf. Proposition \bref{propglob}. This relation re-expressed in the logarithmic form is analogous to that between the global and linearized blocks on the sphere originally proposed  in the 4-point case  \cite{Fitzpatrick:2015zha} and further  developed  in the higher
point case  \cite{Alkalaev:2015fbw}.

Using the Proposition \bref{fkw} we can explicitly find coefficients $\varkappa_n$ in \eqref{lincoefs} as $\varkappa_n =\frac{4^n (2n-1)n}{2}$, cf. first 5 coefficients listed below \eqref{lincoefs}.  Using the integral representation of the global block \eqref{gint}  and the Proposition \bref{fkw} we suggest the closed formula for the linearized classical block
\be
\label{gint2}
f^{lin}(\epsilon,\tilde\epsilon|q)=\int_0^q dx\left( -\frac{\tilde\epsilon}{x}+\sqrt{\frac{\tilde\epsilon^2}{x^2}+\frac{\epsilon^2}{(1-x)^2x}}\,\right)\;.
\ee
From the AdS/CFT perspective, the linearized  block is naturally realized as the tadpole geodesic graph in the thermal AdS space \cite{Alkalaev:2016ptm}. It would be interesting to reproduce this integral formula in the bulk terms thereby proving the correspondence in all orders of the lightness parameter $\delta$.

\section{Conclusion}
\label{sec:con}

The one-point torus block depends on three conformal parameters which particular asymptotic behavior defines different type limiting block functions. In this paper we explicitly described four torus blocks calculated at different scalings of conformal dimensions with respect to the central charge which is either large or infinite. We showed that such semiclassical blocks  are related to each other.  In particular, we found the integral representation of the linearized classical torus block that is important from the AdS/CFT perspective. We showed  that semiclassical torus blocks corresponding to the infinite central charge are associated to differently contracted Virasoro algebra. These are the global, light, and heavy-light blocks.

On the other hand, the (linearized) classical block one defines as large central charge asymptotic expansion of the original Virasoro block can be related to $1/c$ deformations of one of the contracted Virasoro algebras considered in this paper. Indeed, the representation theory of the contracted Lie algebras \cite{Inonu:1953sp} is essentially based on the limiting relation $\varepsilon \nu \sim 1$ between the contraction parameter $\varepsilon \to 0$ and  dimensions  $\nu$ of the non-contracted algebra representations. In other words, the dimension infinitely grows when the contraction parameter goes to zero. In our case, this relation is reproduced by the condition $\Delta/c \sim 1$, see \eqref{epsilon}. We expect that the semiclassical blocks can be entirely reformulated in terms of the representation theory of the contracted Virasoro algebras and their deformations.

The same contraction procedure can be equally applied in the case of the sphere as well as in CFTs on surfaces of any genus. We note also that the Virasoro algebra contractions and associated blocks studied in this paper has much in common with the so-called irregular conformal blocks and their semiclassical limits, see, e.g., \cite{Gaiotto:2012sf,Piatek:2014lma,Rim:2015tsa}. In that case the irregular blocks are associated to the particularly truncated Virasoro algebra and demonstrate the exponential behavior in the limit $\cl$.

Hopefully, further study of the symmetry arguments underlying various semiclassical blocks will improve our understanding of the AdS/CFT duality in the large-$c$ regime. In particular, that will help to clarify how the contracted Virasoro algebras are realized in terms of the bulk geometry and the associated geodesic networks.

\vspace{7mm}

\noindent \textbf{Acknowledgements.} We are grateful to I. Tipunin for useful discussions.  The work of K.A. was supported by the RFBR grant No 14-01-00489. The work of V.R. was performed with the financial support of the Russian Science Foundation (Grant No.14-50-00150).

\appendix

\section{Torus block coefficients}
\label{sec:C}
A few first expansion coefficients of the torus block \eqref{1ptcoef} are given by
\be
\cV_0(\Delta, \tilde\Delta, c)= 1\;,
\qquad
\cV_1(\Delta, \tilde\Delta, c)= 1+\frac{(\Delta - 1)\Delta}{2\tilde \Delta}\;,
\ee
\begin{multline}
\cV_{2}(\Delta, \tilde\Delta, c)=\frac{1}{4 \tilde{\Delta } \left(c-10 \tilde{\Delta }+2 c \tilde{\Delta }+16 \tilde{\Delta
}^2\right)}\times \\
\bigg\{-2 c \Delta +3 c \Delta ^2-2 c \Delta ^3+c \Delta ^4+8 c
\tilde{\Delta }-8 c \Delta  \tilde{\Delta }+56 \Delta ^2 \tilde{\Delta }+8 c \Delta
^2 \tilde{\Delta }-64 \Delta ^3 \tilde{\Delta }+8 \Delta ^4 \tilde{\Delta }-80 \tilde{\Delta }^2+16 c \tilde{\Delta }^2-\\
128 \Delta  \tilde{\Delta
}^2+128 \Delta ^2 \tilde{\Delta }^2+128 \tilde{\Delta }^3 \bigg\}\;,
\end{multline}
\begin{multline}
\cV_{3}(\Delta, \tilde\Delta, c)=\frac{1}{24 \tilde{\Delta } \left(2+c-7
\tilde{\Delta }+c \tilde{\Delta }+3 \tilde{\Delta }^2\right) \left(c-10 \tilde{\Delta }+2 c \tilde{\Delta }+16 \tilde{\Delta }^2\right)}\times \\
\bigg\{-48 c \Delta -24 c^2 \Delta +176 c \Delta ^2+34 c^2 \Delta ^2-312 c \Delta ^3-21 c^2 \Delta ^3+248 c \Delta ^4+13 c^2 \Delta ^4-72 c \Delta
^5-3 c^2 \Delta ^5+\\
8 c \Delta ^6+c^2 \Delta ^6+144 c \tilde{\Delta }+72 c^2 \tilde{\Delta }+240 \Delta  \tilde{\Delta }+12 c \Delta  \tilde{\Delta
}-108 c^2 \Delta  \tilde{\Delta }-668 \Delta ^2 \tilde{\Delta }+890 c \Delta ^2 \tilde{\Delta }+126 c^2 \Delta ^2 \tilde{\Delta }+\\
1794 \Delta ^3
\tilde{\Delta }-1383 c \Delta ^3 \tilde{\Delta }-36 c^2 \Delta ^3 \tilde{\Delta }-1754 \Delta ^4 \tilde{\Delta }+647 c \Delta ^4 \tilde{\Delta }+18
c^2 \Delta ^4 \tilde{\Delta }+414 \Delta ^5 \tilde{\Delta }-177 c \Delta ^5 \tilde{\Delta }-\\
26 \Delta ^6 \tilde{\Delta }+11 c \Delta ^6 \tilde{\Delta
}-1440 \tilde{\Delta }^2-936 c \tilde{\Delta }^2+216 c^2 \tilde{\Delta }^2-1440 \Delta  \tilde{\Delta }^2-696 c \Delta  \tilde{\Delta }^2-96 c^2
\Delta  \tilde{\Delta }^2-2748 \Delta ^2 \tilde{\Delta }^2+\\
1950 c \Delta ^2 \tilde{\Delta }^2+96 c^2 \Delta ^2 \tilde{\Delta }^2+3024 \Delta ^3 \tilde{\Delta
}^2-1644 c \Delta ^3 \tilde{\Delta }^2+1644 \Delta ^4 \tilde{\Delta }^2+390 c \Delta ^4 \tilde{\Delta }^2-504 \Delta ^5 \tilde{\Delta }^2+\\
24 \Delta^6 \tilde{\Delta }^2+7344 \tilde{\Delta }^3-360 c \tilde{\Delta }^3+144 c^2 \tilde{\Delta }^3+6408 \Delta  \tilde{\Delta }^3-1656 c \Delta  \tilde{\Delta
}^3+72 \Delta ^2 \tilde{\Delta }^3+1656 c \Delta ^2 \tilde{\Delta }^3-\\
7488 \Delta ^3 \tilde{\Delta }^3+1008 \Delta ^4 \tilde{\Delta }^3-10224 \tilde{\Delta
}^4+1584 c \tilde{\Delta }^4-5184 \Delta  \tilde{\Delta }^4+5184 \Delta ^2 \tilde{\Delta }^4+3456 \tilde{\Delta }^5\bigg\}\;.
\end{multline}

\section{Proof of Proposition \bref{gl}}
\label{sec:D}

\vspace{-1mm}

\paragraph{Lower levels.} We analyze the $c$-dependence of the first two block coefficients
\eqref{1ptcoef}. The first level coefficient does not depend on $c$  so that $\cV_1 = \cL_1 = \cF_1$, cf. \eqref{lbcoef}. Now, we explicitly calculate the second level coefficient,
\be
\ba{l}
\cV_2 = \dps\frac{1}{\langle \tilde \Delta| \phi_\Delta| \tilde \Delta\rangle}\,\Big[\langle \tilde \Delta, 0,2| \phi_\Delta| 0,2,
\tilde \Delta\rangle B^{0,2|0,2} + \langle \tilde \Delta, 0,2| \phi_\Delta| 1,1, \tilde \Delta\rangle B^{1,1|0,2}
\\
\\
\dps

\hspace{29mm}+\langle \tilde \Delta, 1,1| \phi_\Delta| 0,2, \tilde \Delta\rangle B^{0,2|1,1}+\langle \tilde \Delta, 1,1| \phi_\Delta|
1,1, \tilde \Delta\rangle B^{1,1|1,1}\Big]\;,
\ea
\ee
where $B^{M|N}$ are elements of the inverse Gram matrix, and $| 0,2, \tilde \Delta\rangle = L_{-2} | \tilde \Delta\rangle$ and  $| 1,1, \tilde \Delta\rangle = L_{-1}^2| \tilde \Delta\rangle$. The Gram matrix $B$ and its inverse $B^{-1}$ are given by
\be
\label{38}
B =\left(
 \begin{array}{cc}
 \dps\frac{c}{2}+4 \tilde \Delta  & 6 \tilde \Delta  \\
 6 \tilde \Delta  & 4 \tilde \Delta  (2 \tilde \Delta +1) \\
\end{array}
\right)\;,
\quad
\dps
B^{-1} = \left(
\begin{array}{cc}
 \dps\frac{4 \tilde \Delta +2}{2 \tilde \Delta  c+c+2 \tilde \Delta  (8 \tilde \Delta -5)} & -\dps\frac{3}{2 \tilde \Delta  c+c+2 \tilde
 \Delta  (8 \tilde \Delta -5)} \\
 -\dps\frac{3}{2 \tilde \Delta  c+c+2 \tilde \Delta  (8 \tilde \Delta -5)} & \dps\frac{c+8 \tilde \Delta }{4 \tilde \Delta  (2 \tilde
 \Delta  c+c+2 \tilde \Delta  (8 \tilde \Delta -5))} \\
\end{array}
\right)\;.
\ee
The matrix elements read
\begin{eqnarray}
&&\langle \tilde \Delta, 0,2| \phi_\Delta| 0,2, \tilde \Delta\rangle = \left(4\Delta(\Delta-1)+ (4\tilde \Delta +c/2)\right)\;\langle
\tilde \Delta| \phi_\Delta| \tilde \Delta\rangle\;,\nonumber\\
&&\langle \tilde \Delta, 0,2| \phi_\Delta| 1,1, \tilde \Delta\rangle = \left(2(\Delta-1)\Delta(\Delta+1)+ 6\tilde \Delta\right)\;\langle
\tilde \Delta| \phi_\Delta| \tilde \Delta\rangle\;,\label{39}
\\
&&\langle \tilde \Delta, 1,1| \phi_\Delta| 1,1, \tilde \Delta\rangle = 4 \tilde \Delta  (2 \tilde\Delta +1)\Big[1+\frac{(\Delta -1)
\Delta }{2 \tilde\Delta }+\frac{(\Delta -1) \Delta  (\Delta ^2-\Delta +4 \tilde\Delta)}{4 \tilde \Delta  (2 \tilde\Delta
+1)}\Big]\;\langle \tilde \Delta| \phi_\Delta| \tilde \Delta\rangle\;.\nonumber
\end{eqnarray}
Note that the last matrix element is proportional to  the second coefficient of the global block \eqref{hadasz}. Noting that any block coefficient \eqref{1ptcoef} can be represented as a trace of the matrix product we find that the second level coefficient can be represented as follows
\be
\label{310}
\ba{c}
\cV_2 = \dps\frac{1}{\langle \tilde \Delta| \phi_\Delta| \tilde \Delta\rangle}\,\tr\left(
 \begin{array}{cc}
 \cO(c^{1})  & \cO(c^{0})  \\
 \cO(c^{0})  & \cO(c^{0}) \\
\end{array}
\right)
\left(
 \begin{array}{cc}
 \cO(c^{-1})  & \cO(c^{-1})  \\
 \cO(c^{-1})  & \cO(c^{0}) \\
\end{array}
\right)
\\
\\
\dps
= \langle \tilde \Delta, 0,2| \phi_\Delta| 0,2, \tilde \Delta\rangle B^{0,2|0,2}+ \langle \tilde \Delta, 1,1| \phi_\Delta| 1,1, \tilde \Delta\rangle B^{1,1|1,1} +\cO(c^{-1})\;.
\ea
\ee
In the limit  $c\to \infty$ we find that the block coefficient is given by the sum of diagonal terms while off-diagonal decay as $1/c$. We obtain $\lim_{c\to \infty}\cV_2 \equiv \cL_2 = 1+ \cF_2$. Thus, we have reproduced relation \eqref{LvsF} at $n=0,1,2$. Indeed, $\cL_0 = \cF_0=1$, $\cL_1 = \cF_1$, and $\cL_2 = \cF_2 + \cF_0$.

\vspace{-1mm}

\paragraph{Higher levels.} Now, we analyze the general behavior of the Gram matrix and its inverse in the $\cl$ limit. We show that  the leading  part of the Gram matrix determinant (the Kac determinant) and the leading  part of the product of diagonal elements are the same.

\begin{lemma}
\label{lemma2}

Let $B_{M|N}$ be elements of the Gram matrix $B$. Then,
\be
\label{lemma}
\lim_{c\rightarrow \infty}\frac{\prod_{M}B_{M|M}}{\det B}=1\;,
\ee
where $\det B$ is the Kac determinant.
\end{lemma}

\vspace{-1mm}

\noindent The proof is given in Appendix \bref{sec:A}.

Using the lemma we prove that the expansion coefficients \eqref{1ptcoef} are of order $\cO(1/c^\alpha)$, where $\alpha = 0, 1,2,\ldots\;$, while $\cO(c^0)$ terms come only from the diagonal elements of the inverse Gram matrix $B^{M|N}$ and the primary operator matrix $\langle \tilde \Delta, M| \phi_{\Delta}(z) |N, \tilde \Delta\rangle$, i.e. when $M=N$. Any off-diagonal contributions decay at least as $\cO(1/c)$. It follows that in the $\cl$ limit just diagonal elements contribute. See our discussion surrounding the second level case  \eqref{310}.

We note that the Gram matrix elements and the primary operator matrix have the same behavior with respect to the central charge $c$. By this we mean that the commutation relations of primary operators with any Virasoro generators do not depend on $c$ so that a power-law $c^\beta$, $\beta = 0,1,2,\ldots$ may arise only when commuting two or more Virasoro generators, cf. \eqref{virasoro} and \eqref{primary}. We conclude that estimating powers of the central charge in the block coefficients we can trade the primary operator matrix elements for the Gram matrix elements (cf.  \eqref{38} and \eqref{39}).

We consider  the inverse Gram matrix elements,
\be
\label{inv}
B^{M|N} = \frac{A_{M|N}}{\det B}\;,
\qquad
\det B = \sum_{|M| =|N| = n} (-)^{M+N} B_{M|N} A_{M|N}\;,
\ee
where $A$ is a minor of $B$.  Let $\phi_{M|N}$ be a shorthand notation for the normalized primary operator matrix element $\frac{\langle \tilde \Delta, M| \phi_{\Delta}(z) |N, \tilde \Delta\rangle}{\langle \tilde \Delta| \phi_{\Delta}(z) |\tilde \Delta\rangle}$. Then, the block coefficient \eqref{1ptcoef} is given by
\be
\label{scheme}
\cV_n = \sum_{M}\frac{\phi_{M|M} A_{M|M}}{\det B}+ \sum_{M \neq N}\frac{\phi_{M|N} A_{M|N}}{\det B}\;,
\ee
where we separated diagonal from off-diagonal contributions. We similarly represent the determinant formula in \eqref{inv} as $\det B = \dps \sum_{M} B_{M|M} A_{M|M} + \sum_{M\neq N} (-)^{M+N} B_{M|N} A_{M|N}$. Using \eqref{lemma} we conclude that off-diagonal contribution in the determinant expansion is of less order in $c$ than the diagonal contribution which in the limit $\cl$ is equal to $\dps\prod_{M}B_{M|M}$.

Now, we are in a position to show that the off-diagonal contribution in \eqref{scheme} is of order $\cO(1/c^k)$, $k = 1,2, \ldots$. Comparing $\phi_{M|N} A_{M|N}$ in \eqref{scheme} with the determinant expansion, recalling that $\phi_{M|N}$ has the same $c$-dependence as $B_{M|N}$,  and using the lemma statement \eqref{lemma} we conclude that $\phi_{M|N} A_{M|N}$ is of less order in $c$ than $\det B$ and, therefore, the off-diagonal contribution to the block coefficient is suppressed  at least as $1/c$. On the other hand, the diagonal contribution is always of order $\cO(c^0)$ because $\phi_{M|M} A_{M|M}$ has the same power of $c$ as $\dps\prod_{M}B_{M|M}$.

Finally, we can explicitly compute the diagonal contribution in \eqref{scheme}. Indeed, from the above it follows that in the  $\cl$ limit the Gram matrix can be taken to be diagonal and, therefore, we can easily find its inverse which is also a diagonal matrix. In particular, we find that $B_{M|M} = \langle \tilde \Delta, M| M, \tilde \Delta\rangle  = \gamma_M(c,s) \langle \tilde \Delta| L_1^s L_{-1}^s | \tilde \Delta\rangle$, where $\gamma_M(c,s)$ are some non-zero coefficients. Recalling that $[\phi_\Delta, L_m] = \cO(c^0)$ we similarly find that $\phi_{M|M} = \gamma_M(c,s)\langle \tilde \Delta|L_{1}^s \phi_{\Delta} L^s_{-1}|\tilde \Delta\rangle$. On substituting these expressions into \eqref{1ptcoef} we arrive at the $n$-th level light block coefficient  given by
\be
\label{3.15}
\cL_{n}(\Delta, \tilde\Delta)=\sum_{k=0}^{n}p_1(n-k)\mathcal{F}_k(\Delta, \tilde\Delta)\;,
\ee
where $p_1(n-k)$ denotes the number of partitions of $n-k$ which does not contain $1$ as a part, with the convention that $p_1(0) = 1$, cf. \eqref{LvsF}. Coefficients $p_{1}(n-k)$ count the number of basis ket vectors on the $n$-th level which contain $k$ Virasoro generators $L_{-1}$. Recalling the global block  \eqref{glob} and using \eqref{3.15} and \eqref{generating} we find that
\be
\frac{1-q}{\varphi(q)}\cF(\Delta, \tilde\Delta|q)= p_1(0)+p_1(0)\cF_1(\Delta, \tilde\Delta) q+\big(p_1(0)\cF_2(\Delta, \tilde\Delta)+p_1(2)\cF_0(\Delta, \tilde\Delta)\big)q^2+ ... \;,
\ee
and thereby show the relation between the light and global torus blocks \eqref{lightglob}.

\subsection{Proof of Lemma \bref{lemma}}
\label{sec:A}

\paragraph{Kac determinant.} In  what follows we use the Liouville parametrization of the central charge $c=1+6(b+b^{-1})^2$. Then, the limit $\cl$ can be replaced by $b\to 0$ so that $c = 6/b^2$. The Kac determinant \cite{Kac,Feigin:1981st} in our case is given by
\be
\det B= \prod_{r,s\ge 1}^{rs\le n} ((2r)^{s}s!)^{m(r,s)}(\tilde \de-\de_{r,s})^{p(n-rs)}\;,
\ee
where $p(n)$ is the number of all partitions of  $n$, $\de_{r,s}$ is the Liouville parametrization of the degenerate conformal
dimensions $\de_{r,s}=\frac{(b+b^{-1})^2}{4}-\frac{(rb+sb^{-1})^2}{4}$,  and $m(r,s)=p(n-rs)-p(n-r(s+1))$.

Considering factors with $s = 1$ and $s\geq 2$ separately and taking the $b\to 0$ limit  we find that the $\tilde \Delta-$dependent term
factorizes into the two parts
\be\label{Kac}
\det B=\prod_{k= 1}^{n}\bigg(\tilde \Delta+\frac{k-1}{2}\bigg)^{p(n-k)}\prod_{j\ge 1,i\ge
2}^{ij\le  n}   \bigg(\frac{i^2-1}{4b^2}\bigg)^{p(n-ij)}\prod_{r,s\ge 1}^{rs\le n}((2r)^s s!)^{m(r,s)}\;.
 \ee

\paragraph{Diagonal elements.} We consider diagonal elements of the Gram matrix in the $b \to 0$ limit. In a given diagonal element we take one $L_m$ to the right,

\be
\label{A3}
\ba{c}
  \langle \tilde \Delta |L_{1}^{i_1}...L_{m}^{i_m}L_{-m}^{i_m}...L_{-1}^{i_{1}}| \tilde \Delta\rangle = {\langle \tilde \Delta
  |L_{1}^{i_1}...L_{m}^{i_m-1}[L_{m},L_{-m}^{i_m}...L_{-1}^{i_{1}}]| \tilde \Delta\rangle} =
\\
\\
={\langle \tilde \Delta |L_{1}^{i_1}...L_{m}^{i_m-1}[L_{m},L_{-m}]L_{-m}^{i_m-1}...L_{-1}^{i_{m}}| \tilde \Delta\rangle}+  {\langle
  \tilde \Delta |L_{1}^{i_1}...L_{m}^{i_m-1}L_{-m}[L_{m},L_{-m}^{i_m-1}...L_{-1}^{i_{m}}]| \tilde \Delta\rangle}\;.
\ea
\ee
The leading contribution comes from multiple commutators $\dps{\lim_{b\to 0}[L_{m},L_{-m}]} = m(m^2-1)/(2b^2)$, cf. the central
extension term in \eqref{virasoro}. Commuting all $L_{m}$ to the right we find that  $L_{m}^{i_m}L_{-m}^{i_m}$ contributes with the
combinatorial factor $i_m!$. Finally, we are left only with $L_{-1}^{i_1}$ giving rise to   ${\langle \tilde \Delta|L_{1}^{i_1}L_{-1}^{i_{1}}| \tilde \Delta\rangle}=i_{1}!(2\tilde\de)_{i_1}$.

Now, we describe the $n$-th level basis monomials $L_{-m}^{i_m}...L_{-1}^{i_{1}}$ and their conjugates using partitions of $n$. It follows that the product of diagonal elements can be conveniently represented in terms of restricted partitions as
\be
\label{product}
\prod_{M}B_{M|M}=\prod_{k=1}^{n}\bigg(k!(2\tilde\de)_{k}\bigg)^{p_1(n-k)}\prod_{i=2}^n\bigg(\frac{i^3-i}{2b^2}\bigg)^{\#(i,n)}\;\prod_{r,s\ge2}^{rs\le n}(s!)^{p_r(n-rs)}\;,
\ee
where $p_d(n)$ is the number of partitions of $n$ which does not contain $d$ as a part and $\#(i,n)$ is the number of all parts $i$ in all partitions of $n$. The first product in \eqref{product} accounts for the basis monomials that contain exactly $k$ generators $L_{-1}$ and
$p_1(n-k)$ is the number of partitions of $n$ which contain exactly $k$ units. The second product accounts for  generators $L_{-i}$ with $i\geq 2$ in the set of all partitions. The third product is the  combinatorial factor due to that a partition may contain more than one $L_{-i}$ with $i\geq 2$. To explain this we note that a given partition contains two equal parts provided that $ k =1, ..., n$ can be represented as a product of two natural numbers, $ k = rs$, where $r,s\in \mathbb{N}$. In what follows we express the restricted partitions discussed above in terms of the standard partitions $p(n)$.

\paragraph{Auxiliary relations.} The following relations are useful in practice,
\begin{align}
& \#(i,n)=\sum_{k=1}^{ki\le n}kp_i(n-ki)\;, \label{R1}\\
& p(n) = \sum_{k=0}^{kd\le n}p_d(n-kd)\;, \label{R2}\\
& m(r,s)=p_r(n-rs)\;,\label{R3}\\
& \sum_{k=1}^{ki\le n}p(n-ki) = \sum_{k=1}^{ki\le n}kp_i(n-ki)\;. \label{R4}
\end{align}

To show \eqref{R1} we note that the total number of parts in all partitions of $n$ can be calculated as follows: a number of partitions containing exactly one part $d$ is equal to $p_d(n-d)$,  a number of partitions  containing exactly two parts $d$ is equal to $p_d(n-2d)$, etc. Indeed, fixing a part $d$ and and its number in a given partition $k$ we split the number $n-kd$ into all parts except for $d$ so that there are $p_d(n-kd)$ ways to do it.  Thus, we obtain the total number $p_d(n-d)+2 p_d(n-2d)+... +k p_d(n-k d)\,$, where $k$ is the maximum integer such that $n-kd\geq 0$.

To show \eqref{R2} we consider the generating functions for partitions  $p_d(n-kd)$,
\be
\label{generating}
\frac{1-q^d}{\prod_{k=1}^\infty (1-q^k)}=p_d(0)+p_d(1)q+p_d(2)q^2+p_d(3)q^3 + ... \;,
\ee
see, e.g., \cite{Andrews:1976:TP}, along with the inverse Euler function,
\be
\label{1generating}
\frac{1}{\prod_{k=1}^\infty (1-q^k)}=p(0)+p(1)q+ ... + p(n)q^n + ... \;.
\ee
The function \eqref{1generating} can be represented as a product of two series
\be
\ba{c}\label{2generating}
\dps\frac{1}{\prod_{k=1}^\infty (1-q^k)}=\left(\frac{1}{1-q^d}\right)\left(\frac{1-q^d}{\prod_{k=1}^{\infty}(1-q^k)}\right)=p_d(0)q^0+...+\left(\sum_{k=0}^{kd\le n}p_d(n-kd)\right)q^n+...\;.
\ea
\ee
Comparing the expansion coefficients in \eqref{1generating} and \eqref{2generating} we arrive at the relation  \eqref{R2}.

The relation \eqref{R3} follows from \eqref{R2},
\begin{multline}
m(r,s):=p(n-rs)-p(n-r(s+1))=\\
\big(p_r(n-rs)+p_r(n-rs-r)+...\big)-\big(p_r(n-r(s+1))+p_r(n-r(s+1)-r)+...\big)=\\
p_r(n-rs)\;.
\end{multline}

Finally, to show \eqref{R4} we write down  \eqref{R2} at different arguments,
\be
\ba{c}
\dps
p(n-d)=\sum_{k=0}^{kd\le n-d}p_d(n-d-kd)=\sum_{k=1}^{kd\le n}p_d(n-kd)\;,
\\
\\
\dps
p(n-2d)=\sum_{k=0}^{kd\le n-2d}p_d(n-2d-kd)=\sum_{k=2}^{kd\le n}p_d(n-kd)\;,
\\
\cdots\cdots\cdots\cdots\cdots\cdots\cdots\cdots
\\
p(n-rd)=p_d(n-kd)\;,
\ea
\ee
where $r$ is maximum integer such that $n-rd\ge0$. Summing up all the lines we obtain  \eqref{R2}.

\paragraph{Equivalent representations.} We prove that \eqref{Kac} and \eqref{product} are equal.  On the one hand,
\begin{multline}
\label{F1}
\det B= \prod_{k= 1}^{n}\bigg(\tilde \Delta+\frac{k-1}{2}\bigg)^{p(n-k)}\prod_{j\ge
1,i\ge
2}^{ij\le  n}   \bigg(\frac{i^2-1}{4b^2}\bigg)^{p(n-ij)}\prod_{r,s\ge 1}^{rs\le n}((2r)^s s!)^{m(r,s)}=\\
\prod_{k=
1}^{n}\bigg(\tilde \Delta+\frac{k-1}{2}\bigg)^{p(n-k)}\prod_{i\ge 1,j\ge
2}^{rs\le  n}   \bigg(\frac{i^2-1}{4b^2}\bigg)^{p(n-ij)}\prod_{s= 1}^{ n} 2^{sp_1(n-s)}\prod_{r\ge 2,s\ge 1}^{rs\le n}(2r)^{sp_r(n-rs)}\prod_{r,s\ge 1}^{rs\le n} (s!)^{p_r(n-rs)}=\\
\prod_{k=
1}^{n}\bigg(\tilde \Delta+\frac{k-1}{2}\bigg)^{p(n-k)}\prod_{j\ge 1,i\ge
2}^{ij\le  n}   \bigg(\frac{i^2-1}{4b^2}\bigg)^{p(n-ij)}\prod_{s= 1}^{ n} 2^{p(n-s)}\prod_{r\ge 2,s\ge 1}^{rs\le n}(2r)^{p(n-rs)}\prod_{r,s\ge 1}^{rs\le n} (s!)^{p_r(n-rs)}=\\
\prod_{k= 1}^{n}\bigg(2\tilde \Delta+k-1\bigg)^{p(n-k)}\prod_{j\ge 1,i\ge
2}^{ij\le  n}   \bigg(\frac{i^3-i}{2b^2}\bigg)^{p(n-ij)}\prod_{r,s\ge 1}^{rs\le n} (s!)^{p_r(n-rs)}\;.
\end{multline}
On the other hand,
\begin{multline}
\label{F2}
 \prod_{M}B_{M|M}=\prod_{k=1}^{n}\bigg(k!(2\tilde\de)_{k}\bigg)^{p_1(n-k)}\prod_{i=2}^n\bigg(\frac{i^3-i}{2b^2}\bigg)^{\#(i,n)}\prod_{r,s\ge2}^{rs\le
 n}(s!)^{p_r(n-rs)}=\\ 
 \prod_{k=1}^{n}\bigg(2\tilde\de+k-1\bigg)^{\sum_{i=k}^{n}p_1(n-i)}\prod_{i=2}^n\bigg(\frac{i^3-i}{2b^2}\bigg)^{\sum_{k=1}^{ki\le
 n}kp_i(n-ki)}\prod_{s\ge 1,r\ge2}^{rs\le n}(s!)^{p_r(n-rs)}=\\
 \prod_{k=1}^{n}\bigg(2\tilde\de+k-1\bigg)^{p(n-k)}\prod_{i=2}^n\bigg(\frac{i^3-i}{2b^2}\bigg)^{\sum_{k=1}^{ki\le n}p(n-ki)}\prod_{s\ge
 1,r\ge2}^{rs\le n}(s!)^{p_r(n-rs)}=\\
  \prod_{k=1}^{n}\bigg(2\tilde\de+k-1\bigg)^{p(n-k)}\prod_{i\ge 2, j\ge 1}^{ij\le n}\bigg(\frac{i^3-i}{2b^2}\bigg)^{p(n-ij)}\prod_{r,
  s\ge 1}^{rs\le n}(s!)^{p_r(n-rs)}\;.
\end{multline}
The last lines in \eqref{F1} and \eqref{F2} coincide, and, therefore, the relation \eqref{lemma} holds true.

\section{Recursive representation}
\label{sec:rec}

The recursive representations split  conformal blocks into the limiting function and the singular part with respect to a given conformal parameter which is either the central charge or the exchanged  dimension. The $\tilde\Delta$-recursion representation was proposed in \cite{Hadasz:2009db}. In what follows we carefully derive the $c$-recursive representation.

Using the Liouville parametrization we change
conformal parameters  $(\Delta,\tilde\Delta,c)\rightarrow(\lambda ,\tilde\Delta,b )$ as follows
\be
c(b)=1+6(b+b^{-1})^2\;,
\qquad
\Delta=\frac{(b+b^{-1})^2}{4}-\frac{\lambda^2}{4}\;.
\ee
Conformal blocks have poles at $\tilde \Delta=\Delta_{r,s}$,  which means that the exchanged channels can be degenerate with dimensions
\be
\label{deltars}
\Delta_{rs}=\frac{(b+b^{-1})^2}{4}-\frac{(rb+sb^{-1})^2}{4}\;.
\ee
The $\tilde\Delta$-recursive representation of 1-point torus blocks is given by \cite{Hadasz:2009db}
\be
\label{d-recursion}
\cV_n(\lambda ,\tilde\Delta,b )=\mathcal{H}_n(\lambda , b )+\sum_{r,s\ge1}^{rs\le n}\frac{A_{rs}(b) P_{rs}(\lambda,b)}{\tilde \Delta-\Delta_{rs}(b)}\cV_{n-rs}(\lambda, \Delta_{rs}+rs, b)\;,
\ee
where $\mathcal{H}_n(\lambda , b )$ are the expansion coefficients of the $\tilde\Delta \rightarrow \infty$ limiting conformal block given by the inverse Euler function, cf. \eqref{1generating}, and
\be
A_{rs}(b)=\frac{1}{2}\prod_{\substack{p=1-r\\(p,q)\ne(0,0)\ne (r,s)}}^{r}\prod_{\substack{q=1-s }}^{s}(pb+qb^{-1})^{-1}\;,
\ee
\begin{multline}
 P_{rs}(\lambda,b)=\prod_{k=1, k=1 (mod 2)}^{2r-1}\prod_{l=1,l=1(mod
 2)}^{2s-1}\bigg(\frac{\lambda+kb+lb^{-1}}{2}\bigg)\bigg(\frac{\lambda-kb+lb^{-1}}{2}\bigg)\times\\
 \bigg(\frac{\lambda+kb-lb^{-1}}{2}\bigg)\bigg(\frac{\lambda-kb-lb^{-1}}{2}\bigg)\;.
\end{multline}

Using the recursive formula \eqref{d-recursion} we can elaborate the other recursive representation, where poles are given on the $c$-plane rather than $\tilde\Delta$-plane. Consequently, the leading asymptotic in this case is the $\cl$ limiting  block which is the light block \eqref{lbcoef}. To elaborate the $c$-recursive representation we first re-parameterize the conformal block poles given by $\tilde\Delta - \Delta_{rs}(b)=0$, cf. \eqref{d-recursion} and \eqref{deltars}. Considering this condition as the equation on the $b$-plane
\be
\tilde \Delta =\frac{(b_{r,s}+b_{r,s}^{-1})^2}{4}-\frac{(rb_{r,s}+sb_{r,s}^{-1})}{4}\;,
\ee
we find that poles arise when the central charge $b$ takes particular ''degenerate'' values,
\be
b_{r,s}^2(\tilde\Delta)=\frac{1}{1-r^2}\bigg(2\tilde\Delta+rs-1+\sqrt{(r-s)^2+4(rs-1)\tilde\Delta+4\tilde\Delta^2}\bigg)\;.
\ee
It follows that \eqref{d-recursion} can be equivalently represented as
\be
\label{vtornik}
    \cV_n(\lambda ,\tilde\Delta,b ) = \cL_n(\lambda, \tilde\Delta) + \sum_{r\ge2,s\ge1}^{rs\le n}\bigg(-\frac{\partial b_{r,s}}{\partial
    \tilde \Delta}\bigg)\frac{A_{rs}(b_{r,s}) P_{rs}(\lambda,b_{r,s})}{b-b_{rs}(\tilde\Delta)}\cV_{n-rs}(\lambda, \tilde\Delta+rs, b)\;,
\ee
where the leading term is the light block \eqref{lbcoef}. The Jacobian prefactor follows from comparing the residua,
\be
\ba{l}
\dps
res_{b_{r,s}}\cV_n(\lambda ,\tilde\Delta,b )=\frac{1}{2\pi i}\oint_{b_{r,s}}\cV_n(\lambda ,\tilde\Delta,b )db =
\\
\\
\dps
\hspace{30mm}=\frac{1}{2\pi i}\oint_{\Delta_{r,s}}\cV_n(\lambda ,\tilde\Delta,b
)\bigg(\frac{\partial c}{\partial \tilde\Delta}\bigg)d\tilde\Delta=\bigg(\frac{\partial b_{r,s}}{\partial
\tilde\Delta}\bigg)res_{\Delta_{r,s}}\cV_n(\lambda ,\tilde\Delta,b)\;.
\ea
\ee
The minus sign of the Jacobian in \eqref{vtornik} is due to  interchanging terms of the pole part when coming from the $\tilde\Delta$-plane to the $b$-plane.

Finally, using the standard parametrization
\be
\begin{aligned}
&c_{r,s}=1+6(b_{r,s}+b_{r,s}^{-1})^2\;,\\
&\lambda^2=(b+b^{-1})^2-4\Delta\;,\\
&b^2=\frac{c-13+\sqrt{25-26c+c^2}}{12}\;,
 \end{aligned}
\ee
the recursive formula \eqref{vtornik} can be rewritten as
\be
\label{torusrecApp}
    \cV_n(\Delta,\tilde\Delta,c ) = \cL_n(\Delta, \tilde\Delta) + \sum_{r\ge2,s\ge1}^{rs\le n}\bigg(-\frac{\partial c_{r,s}}{\partial
    \tilde \Delta}\bigg)\frac{A_{rs}(c_{r,s}) P_{rs}(\tilde\Delta,c_{r,s})}{c-c_{rs}(\tilde\Delta)}\cV_{n-rs}(\Delta, \tilde\Delta+rs,c)\;,
\ee
where the Jacobian factor is given by
\be
\frac{\partial c_{r,s}}{\partial \tilde\Delta}=\frac{12}{b_{r,s}^2}\frac{1-b_{r,s}^4}{(b_{r,s}^2(r^2-1)+2\tilde\Delta+rs-1)}\;.
\ee
Using the recursive representation \eqref{torusrecApp} we explicitly calculated the first few  block coefficients $\cV_k(\Delta,\tilde\Delta,c)$, $k=0,1,2,3$ and found that they reproduce the standard expressions collected in Appendix \bref{sec:C}.



\begin{thebibliography}{10}

\bibitem{Brown:1986nw}
J.~D. Brown and M.~Henneaux, \emph{{Central Charges in the Canonical
  Realization of Asymptotic Symmetries: An Example from Three-Dimensional
  Gravity}},
  \href{http://dx.doi.org/10.1007/BF01211590}{\emph{Commun.Math.Phys.} {\bf
  104} (1986) 207--226}.

\bibitem{Hartman:2013mia}
T.~Hartman, \emph{{Entanglement Entropy at Large Central Charge}},
  \href{http://arxiv.org/abs/1303.6955}{{\tt 1303.6955}}.

\bibitem{Fitzpatrick:2015zha}
A.~L. Fitzpatrick, J.~Kaplan and M.~T. Walters, \emph{{Virasoro Conformal
  Blocks and Thermality from Classical Background Fields}},
  \href{http://dx.doi.org/10.1007/JHEP11(2015)200}{\emph{JHEP} {\bf 11} (2015)
  200}, [\href{http://arxiv.org/abs/1501.05315}{{\tt 1501.05315}}].

\bibitem{Hijano:2015rla}
E.~Hijano, P.~Kraus and R.~Snively, \emph{{Worldline approach to semi-classical
  conformal blocks}},
  \href{http://dx.doi.org/10.1007/JHEP07(2015)131}{\emph{JHEP} {\bf 07} (2015)
  131}, [\href{http://arxiv.org/abs/1501.02260}{{\tt 1501.02260}}].

\bibitem{Alkalaev:2015wia}
K.~B. Alkalaev and V.~A. Belavin, \emph{{Classical conformal blocks via AdS/CFT
  correspondence}},
  \href{http://dx.doi.org/10.1007/JHEP08(2015)049}{\emph{JHEP} {\bf 08} (2015)
  049}, [\href{http://arxiv.org/abs/1504.05943}{{\tt 1504.05943}}].

\bibitem{Hijano:2015qja}
E.~Hijano, P.~Kraus, E.~Perlmutter and R.~Snively, \emph{{Semiclassical
  Virasoro blocks from AdS$_{3}$ gravity}},
  \href{http://dx.doi.org/10.1007/JHEP12(2015)077}{\emph{JHEP} {\bf 12} (2015)
  077}, [\href{http://arxiv.org/abs/1508.04987}{{\tt 1508.04987}}].

\bibitem{Banerjee:2016qca}
P.~Banerjee, S.~Datta and R.~Sinha, \emph{{Higher-point conformal blocks and
  entanglement entropy in heavy states}},
  \href{http://dx.doi.org/10.1007/JHEP05(2016)127}{\emph{JHEP} {\bf 05} (2016)
  127}, [\href{http://arxiv.org/abs/1601.06794}{{\tt 1601.06794}}].

\bibitem{Chen:2016dfb}
B.~Chen, J.-q. Wu and J.-j. Zhang, \emph{{Holographic Description of 2D
  Conformal Block in Semi-classical Limit}},
  \href{http://dx.doi.org/10.1007/JHEP10(2016)110}{\emph{JHEP} {\bf 10} (2016)
  110}, [\href{http://arxiv.org/abs/1609.00801}{{\tt 1609.00801}}].

\bibitem{Alkalaev:2016rjl}
K.~B. Alkalaev, \emph{{Many-point classical conformal blocks and geodesic
  networks on the hyperbolic plane}},
  \href{http://arxiv.org/abs/1610.06717}{{\tt 1610.06717}}.

\bibitem{Fitzpatrick:2016mjq}
A.~L. Fitzpatrick and J.~Kaplan, \emph{{On the Late-Time Behavior of Virasoro
  Blocks and a Classification of Semiclassical Saddles}},
  \href{http://arxiv.org/abs/1609.07153}{{\tt 1609.07153}}.

\bibitem{Hulik:2016ifr}
O.~Hul\'{i}k, T.~Proch\'{a}zka and J.~Raeymaekers, \emph{{Multi-centered
  AdS$_3$ solutions from Virasoro conformal blocks}},
  \href{http://arxiv.org/abs/1612.03879}{{\tt 1612.03879}}.

\bibitem{Cardy:1986ie}
J.~L. Cardy, \emph{{Operator Content of Two-Dimensional Conformally Invariant
  Theories}}, \href{http://dx.doi.org/10.1016/0550-3213(86)90552-3}{\emph{Nucl.
  Phys.} {\bf B270} (1986) 186--204}.

\bibitem{Fateev:2009me}
V.~A. Fateev, A.~V. Litvinov, A.~Neveu and E.~Onofri, \emph{{Differential
  equation for four-point correlation function in Liouville field theory and
  elliptic four-point conformal blocks}},
  \href{http://dx.doi.org/10.1088/1751-8113/42/30/304011}{\emph{J. Phys.} {\bf
  A42} (2009) 304011}, [\href{http://arxiv.org/abs/0902.1331}{{\tt
  0902.1331}}].

\bibitem{Poghossian:2009mk}
R.~Poghossian, \emph{{Recursion relations in CFT and N=2 SYM theory}},
  \href{http://dx.doi.org/10.1088/1126-6708/2009/12/038}{\emph{JHEP} {\bf 12}
  (2009) 038}, [\href{http://arxiv.org/abs/0909.3412}{{\tt 0909.3412}}].

\bibitem{Hadasz:2009db}
L.~Hadasz, Z.~Jaskolski and P.~Suchanek, \emph{{Recursive representation of the
  torus 1-point conformal block}},
  \href{http://dx.doi.org/10.1007/JHEP01(2010)063}{\emph{JHEP} {\bf 01} (2010)
  063}, [\href{http://arxiv.org/abs/0911.2353}{{\tt 0911.2353}}].

\bibitem{Menotti:2010en}
P.~Menotti, \emph{{Riemann-Hilbert treatment of Liouville theory on the
  torus}}, \href{http://dx.doi.org/10.1088/1751-8113/44/11/115403}{\emph{J.
  Phys.} {\bf A44} (2011) 115403}, [\href{http://arxiv.org/abs/1010.4946}{{\tt
  1010.4946}}].

\bibitem{Fitzpatrick:2014vua}
A.~L. Fitzpatrick, J.~Kaplan and M.~T. Walters, \emph{{Universality of
  Long-Distance AdS Physics from the CFT Bootstrap}},
  \href{http://dx.doi.org/10.1007/JHEP08(2014)145}{\emph{JHEP} {\bf 1408}
  (2014) 145}, [\href{http://arxiv.org/abs/1403.6829}{{\tt 1403.6829}}].

\bibitem{Alkalaev:2015lca}
K.~B. Alkalaev and V.~A. Belavin, \emph{{Monodromic vs geodesic computation of
  Virasoro classical conformal blocks}},
  \href{http://dx.doi.org/10.1016/j.nuclphysb.2016.01.019}{\emph{Nucl. Phys.}
  {\bf B904} (2016) 367--385}, [\href{http://arxiv.org/abs/1510.06685}{{\tt
  1510.06685}}].

\bibitem{Datta:2014zpa}
S.~Datta, J.~R. David and S.~P. Kumar, \emph{{Conformal perturbation theory and
  higher spin entanglement entropy on the torus}},
  \href{http://dx.doi.org/10.1007/JHEP04(2015)041}{\emph{JHEP} {\bf 04} (2015)
  041}, [\href{http://arxiv.org/abs/1412.3946}{{\tt 1412.3946}}].

\bibitem{Rangamani:2016dms}
M.~Rangamani and T.~Takayanagi, \emph{{Holographic Entanglement Entropy}},
  \href{http://arxiv.org/abs/1609.01287}{{\tt 1609.01287}}.

\bibitem{Alkalaev:2016ptm}
K.~B. Alkalaev and V.~A. Belavin, \emph{{Holographic interpretation of 1-point
  toroidal block in the semiclassical limit}},
  \href{http://dx.doi.org/10.1007/JHEP06(2016)183}{\emph{JHEP} {\bf 06} (2016)
  183}, [\href{http://arxiv.org/abs/1603.08440}{{\tt 1603.08440}}].

\bibitem{Alkalaev:2015fbw}
K.~B. Alkalaev and V.~A. Belavin, \emph{{From global to heavy-light: 5-point
  conformal blocks}},
  \href{http://dx.doi.org/10.1007/JHEP03(2016)184}{\emph{JHEP} {\bf 03} (2016)
  184}, [\href{http://arxiv.org/abs/1512.07627}{{\tt 1512.07627}}].

\bibitem{Dolan:2011dv}
F.~Dolan and H.~Osborn, \emph{{Conformal Partial Waves: Further Mathematical
  Results}},  \href{http://arxiv.org/abs/1108.6194}{{\tt 1108.6194}}.

\bibitem{Zamolodchikov:1987ie}
A.~Zamolodchikov, \emph{{Conformal Symmetry in Two-dimensional Space: Recursion
  Representation of the Conformal Block}}, {\emph{Teor.Mat.Fiz.} {\bf 73}
  (1987) 103--110}.

\bibitem{Bhatta:2016hpz}
A.~Bhatta, P.~Raman and N.~V. Suryanarayana, \emph{{Holographic Conformal
  Partial Waves as Gravitational Open Wilson Networks}},
  \href{http://dx.doi.org/10.1007/JHEP06(2016)119}{\emph{JHEP} {\bf 06} (2016)
  119}, [\href{http://arxiv.org/abs/1602.02962}{{\tt 1602.02962}}].

\bibitem{Fateev:2011qa}
V.~Fateev and S.~Ribault, \emph{{The Large central charge limit of conformal
  blocks}}, \href{http://dx.doi.org/10.1007/JHEP02(2012)001}{\emph{JHEP} {\bf
  02} (2012) 001}, [\href{http://arxiv.org/abs/1109.6764}{{\tt 1109.6764}}].

\bibitem{Poghosyan:2016lya}
H.~Poghosyan, R.~Poghossian and G.~Sarkissian, \emph{{The light asymptotic
  limit of conformal blocks in Toda field theory}},
  \href{http://dx.doi.org/10.1007/JHEP05(2016)087}{\emph{JHEP} {\bf 05} (2016)
  087}, [\href{http://arxiv.org/abs/1602.04829}{{\tt 1602.04829}}].

\bibitem{Kraus:2016nwo}
P.~Kraus and A.~Maloney, \emph{{A Cardy Formula for Three-Point Coefficients:
  How the Black Hole Got its Spots}},
  \href{http://arxiv.org/abs/1608.03284}{{\tt 1608.03284}}.

\bibitem{Inonu:1953sp}
E.~Inonu and E.~P. Wigner, \emph{{On the Contraction of groups and their
  represenations}}, \href{http://dx.doi.org/10.1073/pnas.39.6.510}{\emph{Proc.
  Nat. Acad. Sci.} {\bf 39} (1953) 510--524}.

\bibitem{Barut:1970qf}
A.~O. Barut and L.~Girardello, \emph{{New 'coherent' states associated with
  noncompact groups}},
  \href{http://dx.doi.org/10.1007/BF01646483}{\emph{Commun. Math. Phys.} {\bf
  21} (1971) 41--55}.

\bibitem{Zamolodchikov1986}
A.~Zamolodchikov, \emph{{Two-dimensional conformal symmetry and critical
  four-spin correlation functions in the Ashkin-Teller model}}, {\emph{Zh.
  Eksp. Teor. Fiz.} {\bf 90} (1986) 1808--1818}.

\bibitem{Piatek:2013ifa}
M.~Piatek, \emph{{Classical torus conformal block, $N = 2^*$ twisted
  superpotential and the accessory parameter of Lame equation}},
  \href{http://dx.doi.org/10.1007/JHEP03(2014)124}{\emph{JHEP} {\bf 03} (2014)
  124}, [\href{http://arxiv.org/abs/1309.7672}{{\tt 1309.7672}}].

\bibitem{Gaiotto:2012sf}
D.~Gaiotto and J.~Teschner, \emph{{Irregular singularities in Liouville theory
  and Argyres-Douglas type gauge theories, I}},
  \href{http://dx.doi.org/10.1007/JHEP12(2012)050}{\emph{JHEP} {\bf 12} (2012)
  050}, [\href{http://arxiv.org/abs/1203.1052}{{\tt 1203.1052}}].

\bibitem{Piatek:2014lma}
M.~Piatek and A.~R. Pietrykowski, \emph{{Classical irregular block, $
  \mathcal{N} $ = 2 pure gauge theory and Mathieu equation}},
  \href{http://dx.doi.org/10.1007/JHEP12(2014)032}{\emph{JHEP} {\bf 12} (2014)
  032}, [\href{http://arxiv.org/abs/1407.0305}{{\tt 1407.0305}}].

\bibitem{Rim:2015tsa}
C.~Rim and H.~Zhang, \emph{{Classical Virasoro irregular conformal block}},
  \href{http://dx.doi.org/10.1007/JHEP07(2015)163}{\emph{JHEP} {\bf 07} (2015)
  163}, [\href{http://arxiv.org/abs/1504.07910}{{\tt 1504.07910}}].

\bibitem{Kac}
V.~G. Kac, \emph{{Contravariant form for infinite dimensional Lie algebras and
  superalgebras }}, {\emph{Springer Verlag, Berlin} {\bf 94} (1979) }.

\bibitem{Feigin:1981st}
B.~L. Feigin and D.~B. Fuks, \emph{{Invariant skew symmetric differential
  operators on the line and verma modules over the Virasoro algebra}},
  \href{http://dx.doi.org/10.1007/BF01081626}{\emph{Funct. Anal. Appl.} {\bf
  16} (1982) 114--126}.

\bibitem{Andrews:1976:TP}
G.~E. Andrews, \emph{The Theory of Partitions}, vol.~2 of \emph{Encyclopedia of
  Mathematics and its Applications}.
\newblock Addison-Wesley Publishing Co., Reading, MA-London-Amsterdam, 1976.

\end{thebibliography}

\providecommand{\href}[2]{#2}\begingroup\raggedright\endgroup

\end{document}